%% file: make_astro.tex
\documentclass{mpe_report}

\usepackage{psfig,graphicx,epsfig}
\usepackage{color}
\usepackage{amsmath,amssymb,epic,eepic,array}

\begin{document}

\pagenumbering{arabic}
\setcounter{page}{205}

 \renewcommand{\FirstPageOfPaper }{205}\renewcommand{\LastPageOfPaper }{207}\include{./mpe_report_tanaka}            \clearpage

\end{document}

%% file: mpe_report_tanaka.tex
\title{GEOTAIL observation of SGR 1900+14 giant flare on 27 August 1998}
\authorrunning{Y. T. Tanaka et al.}
\titlerunning{GEOTAIL observation of SGR 1900+14 giant flare}
\author{Yasuyuki T. Tanaka\inst{1}, Toshio Terasawa\inst{2}, 
Nobuyuki Kawai\inst{2} and Ichiro Yoshikawa\inst{1}}  
\institute{Department of Earth and Planetary Science,
  University of Tokyo, 7-3-1 Hongo, Bunkyo-ku, Tokyo 113-0033, Japan
\and 
  Department of Physics,
  Tokyo Institute of Technology, Tokyo 152-8551, Japan}
\maketitle

\begin{abstract}
The soft gamma repeater (SGR) 1900+14 emitted the giant flare on 27
August 1998. Most gamma-ray detectors saturated during the initial spike 
of the giant flare because of the intense flux. However the plasma particle
detector onboard GEOTAIL observed the first 300 ms time profile 
with a time resolution of 5.577 ms and the initial spike of the giant flare
was first resolved. The time profile shows some similarities to that of the
SGR 1806-20 giant flare in 2004: the clear exponential decay and 
the small hump in the decay phase around 300 or 400 ms.  
\end{abstract}

\section{Introduction}
Soft gamma repeaters (SGRs) are young neutron stars
emitting short and energetic bursts of photons
in soft gamma ray energies.
In addition, SGRs occasionally provide giant flares,
whose energy amounts to 10$^3$-10$^5$ times of
those of repeated bursts.
The first giant flare was discovered
on 5 March 1979 as a sudden increase of soft gamma ray photon fluxes
from SGR 0525-66 in the Large Magellanic Clouds,
a galaxy neighboring to our Galaxy (Mazets et al. 1979).
Since then two giant flares occurred within our Galaxy
on 27 August 1998 and 27 December 2004, the latest of which was stronger
by a factor of hundreds than preceding ones.
While practically all gamma-ray detectors on any satellites 
were saturated during the first $\sim$500 ms interval after the onset
of the 2004 giant flare
(e.g., Hurley et al. 2005, Palmer et al. 2005, Mazets et al. 2005, Mereghetti et al. 2005), 
a few particle detectors were not saturated and provided
important information on the initial very intense spike
(Terasawa et al. 2005, Schwartz et al. 2005).
From the plasma particle detectors on the GEOTAIL spacecraft
the peak photon energy flux (integrated above 50 keV) 
was estimated to be 
the order of 10$^7$ photons sec$^{-1}$ cm$^{-2}$,
the peak energy flux 
$\sim$20 erg sec$^{-1}$cm$^{-2}$
(Terasawa et al. 2005).
That this energy flux was by a factor $>\sim$1000 stronger
than those from largest solar flares
is surprising if we notice that 
the estimated distance to the source of this giant flare 
(SGR1806-20) is 15 kpc,
namely 3$\times 10^{9}$ times farther than the sun.

The ``magnetar" model 
(Duncan and Thompson 1992, Thompson and Duncan 1995)
is generally accepted
to explain the nature of SGRs,
where neutron stars having ultrastrong magnetic field
of the order of 10$^{14}-$10$^{15}$ G
eventually release the magnetic energy to keep repeating soft gamma activity
as well as to cause giant flares.
It is noted that in spite of its success in SGR energetics
the magnetar model still includes hypothetical parts:
For example, the magnetic reconnection process in magnetars' magnetospheres
 is invoked to explain the energy conversion from magnetic fields
 to relativistic pair plasmas at the onset of bursts/giant flares.
Where and how such reconnection process occurs is yet to be studied
both theoretically and observationally.
We expect the data of the initial phase of giant flares should
play an essential role in such studies.
                 
\section{Instrumentation and Calibration}
The Low Energy Particle (LEP) experiment (Mukai et al. 1994) 
onboard GEOTAIL consists of two nested sets of quadspherical
electrostatic analyzers with seven microchannel plates (MCPs)
installed as ion detectors and seven channel electron multipliers (CEMs)
as electron detectors. We can perform gamma-ray observation using the MCPs.

Because the MCPs onboard GEOTAIL are not designed to perform 
gamma-ray observations, we must calibrate them as
the gamma-ray detectors. First we have to conduct Monte Carlo
simulations in order to examine the contaminations of photo-electrons,
compton-electrons, and characteristic X-rays as well as the effect of the scattering with
the satellite body. Next, we have to
measure the detection efficiency of the MCP for 
gamma-rays in the laboratory. 

\begin{figure}
\centerline{\psfig{file=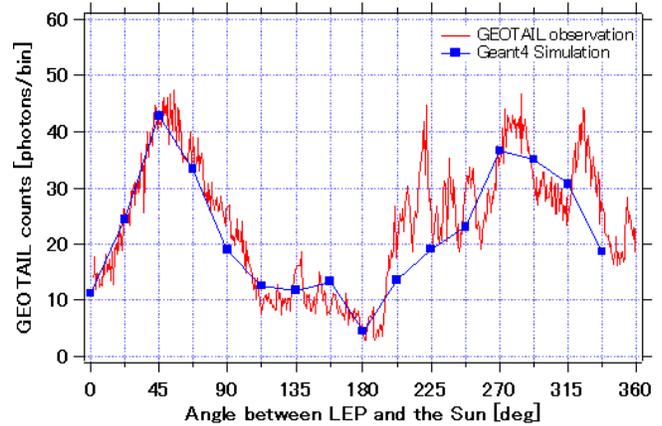,width=8.8cm,clip=} }
\caption{The modulation profile of gamma-rays detected with the plasma particle detectors
onboard GEOTAIL. The horizontal axis shows the angle between the particle detectors and
the sun. The modulation results from the combined effect of the difference of 
the quantity of material along the path to the detectors and the effective area.
\label{spinmodulation}}
\end{figure}

We have constructed the mass model of GEOTAIL based on the 
Geant4 (Agostinelli et al. 2003). 
The validity
of this mass model is confirmed as follows.
When X-class large solar flares occurred, 
the background noise counts of the MCPs
increased predominantly. These were due to the gamma-rays which were emitted 
from the solar flares and came into the detectors directly. 
Because GEOTAIL is a spin-stabilized satellite,
the detected gamma-rays were modulated as the quantity of 
material along the path to the detector differed. 
The modulation profile of gamma-rays during the peak time of the X5.3 solar
flare occurred on 25 August 2001 are shown in Fig.
\ref{spinmodulation} with red line. 
The blue line in 
Fig. \ref{spinmodulation} shows the simulation results. This is obtained
by irradiating gamma-rays to the mass model 
from various azimuthal angles (the incident 
polar angle is fixed as 90 degree because the orbit of GEOTAIL is almost
in the ecliptic plane). Note that the spectrum of the solar flare 
was observed by YOHKOH satellite so that we can use it as the incident 
gamma-ray spectrum.
We can well reproduce the modulation so that we can confirm the validity 
of the mass model. 

We irradiated the numerous gamma-rays whose spectrum was $kT$=240 keV optically
thin thermal bremsstrahlung (Hurley et al. 1999) 
to the mass model of GEOTAIL. 
Because we know both the spin axis of GEOTAIL and the spin phase, 
we can exactly determine the direction of gamma-rays.
From the simulation results
we found that the contaminations of photo-electrons, compton-electrons, and
characteristic X-rays are less than 1 \% compared to the primary gamma-rays.

Next, we prepared the MCP
of the same geometrical size and constituents as was equipped on GEOTAIL and 
measured the
quantum detection efficiency for gamma-rays. In order to examine the energy 
dependency of the efficiency, we used two gamma sources: Cs137 
(662 keV) and Am241 (60 keV). Furthermore, in order to examine the angular
dependency, we irradiated the gamma-rays whose incident angle to the detector
is 0 degree or 180 degree. Preliminary results show that the efficiency is
about 1\% and depends to some extent on the energy and the incident angle of 
gamma-rays.
Note that fine tunings of this experiment is now under way, and
the details of this experiment will be reported elsewhere (Tanaka et al.
in preparation).

It should be noted that the time resolution of this observation 
is 5.577 ms, which is a little different compared to that of the observation
of SGR 1806-20 giant flare in 2004 (Terasawa et al. 2005). This is because
the time resolution is determined by dividing the spin period and the
spin period changes slightly year by year. 

\begin{figure}
\centerline{\psfig{file=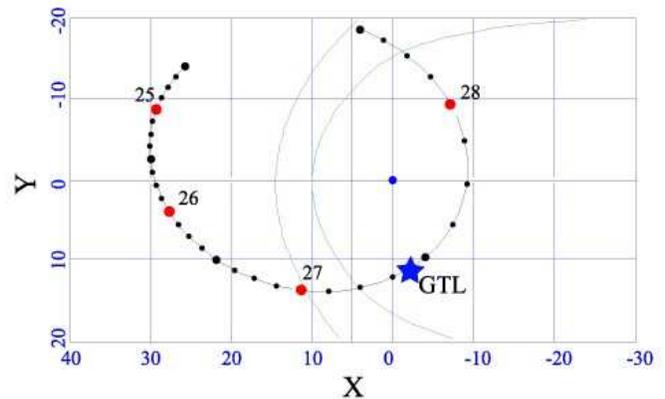,width=8.8cm,clip=} }
\caption{The orbit of GEOTAIL around 27 August 1998. 
The blue star shows the position of
GEOTAIL when the SGR 1900+14 giant flare occurred. Also drawn are the nominal
positions of the earth's bow shock and magnetopause.
\label{orbit}}
\end{figure}

\section{The initial spike of the SGR 1900+14 giant flare on 27 August 1998}
SGR 1900+14 giant flare occurred at 10:22 UT on 27 August 1998.
Fig. \ref{orbit} shows the orbit of GEOTAIL around 27 August 1998.
The position of GEOTAIL at the time when the giant flare occurred is shown
with the blue star and the plasma particle detectors onboard GEOTAIL was 
observing the magnetospheric plasmas. 

The flux of SGR 1900+14 
giant flare was so intense that most of the gamma-ray detectors could not
observe the peak profile of the giant flare because of the saturation effects
or pulse pile-up problems: in fact,
the gamma-ray detectors onboard Ulysses and Konus-Wind
saturated during the initial spike and could determine only the
lower limits of the peak flux intensity and fluence 
(Hurley et al. 1999, Mazets et al. 1999).
On the other hand, we have found that 
the initial time profile was obtained by the plasma particle detector onboard GEOTAIL.
This is because the effective area of the detector is very small and 
the detection efficiency is very low compared to the common gamma-ray detectors on
any satellites.

\begin{figure}
\centerline{\psfig{file=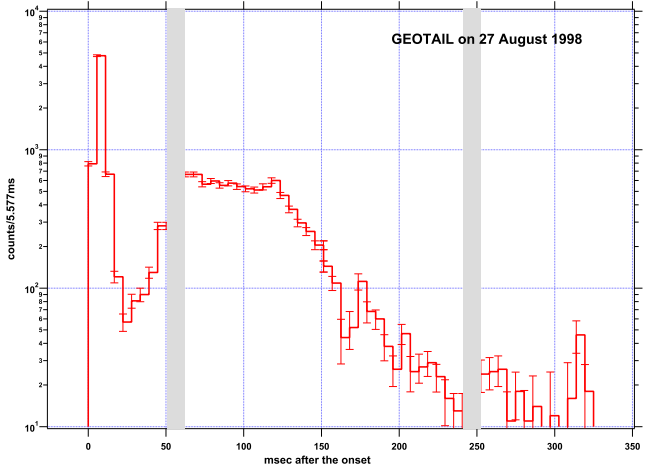,width=8.8cm,clip=} }
\caption{The above 50 keV time profile of the SGR 1900+14 giant flare on 
27 August 1998, corrected for dead-time effects. 
Shaded bars show the instrumental data gaps.
\label{lightcurve}}
\end{figure}

Fig. \ref{lightcurve} shows the first 300 ms profile of the giant flare
with the time resolution of 5.577 ms after the dead time correction.
Note that the time profile is preliminary.
The energy range of the time profile is above $\sim$50 keV, which is confirmed
from the Monte Carlo simulations.
The shaded bars indicate the operational data gaps.
The onset time ($t$=0) corresponded to 10 h 22 min 15.47 s UT.
After the onset the photon counts reached a very sharp peak at $t$=5.58 ms.
Following the peak, there was a dip during the interval of $t$=15$-$45 ms. After the dip, 
the photon counts again increased and reached the flat-top second peak during 60$-$120 ms.
Then the counts decayed exponentially and the decay time was calculated as $\sim$22 ms
using the time profile during the $t$=120$-$240 ms.
Around 310 ms, the small hump was seen in GEOTAIL data
and this was also observed with Konus-Wind (see Fig.6 of Mazets et al. 1999).

The exponential decay and the small hump may be the common features
of the time profiles of the initial spikes. These structures were also 
observed in the time profile of SGR 1806-20 giant flare: the decay time
was $\sim$66 ms and the small hump was observed for $t$=402-451 ms
(Terasawa et al. 2005). These timescales are a little different from those observed
during the SGR 1900+14 giant flare in 1998. The physical meanings of this difference
are unknown so far.

Because the particle detectors onboard GEOTAIL cannot
observe the spectrum, we have to assume the spectrum of the initial spike
in order to estimate the total energy of the giant flare.
We took the $kT$=240 keV optically thin thermal spectrum 
from Hurley et al. (1999) and 
from the preliminary analyses found that the total emitted energy
was the order of the 10$^{44}$ erg. Here we assumed that
the distance to SGR 1900+14 is 10 kpc.  
More detailed analyses are now under way.
 
\section{Conclusions}
In this report we present the GEOTAIL observation of SGR 1900+14 giant flare
on 27 August 1998. The initial spike of the giant flare was first resolved and
the profile consisted of five segments: the main spike, the deep dip, the flat-top 
second peak, the exponential decay and the small hump. 
The similar exponential decay and the small hump was also observed in the SGR 
1806-20 giant flare in 2004 (Terasawa et al. 2005, Palmer et al. 2005). 
Now the experiments which measure the detection efficiency of the plasma particle 
detector equipped with GEOTAIL for gamma-rays are under way.
More detailed results of our analyses of the SGR 1900+14 giant flare in 1998 will be 
reported elsewhere (Tanaka et al. in preparation).

\vskip 0.4cm

\begin{acknowledgements}
We gratefully acknowledge GEOTAIL team for valuable comments and discussions.
\end{acknowledgements}
   
